# EFFECT OF ELASTICITY OF SHAFTS, BEARINGS, CASING AND COUPLINGS ON THE CRITICAL ROTATIONAL SPEEDS OF A GEARBOX.

Emmanuel RIGAUD and Jean SABOT.
Laboratoire de Tribologie et Dynamique des Systèmes. URA N°855. Ecole Centrale de Lyon.
36, avenue Guy de Collongue. B.P. 163. 69131 Ecully Cedex. FRANCE.

***Abstract :*** *The aim of this study is to analyse the influence of the mechanical characteristics of the set of components on the critical rotational speeds of a gearbox. The case of a gearbox fitted out with a helical gear pair was considered. The shafts and the casing were discretised using the finite element method. The elastic coupling between the toothed wheels was characterised by a 12 x12 stiffness matrix. The bearings were modelled using radial, axial and rotational stiffness elements. The calculation of the vibration response induced by the static transmission error showed that the highest dynamic mesh forces correspond to a resonant excitation of modes which have a high potential energy associated with the mesh stiffness. The numerical simulations performed showed that a realistic prediction of the critical rotational speeds should take account of all the components of the gearbox.*

## 1- INTRODUCTION

In the course of designing a geared system, it is important to be able to predict those rotational speeds which will lead to the highest dynamic mesh forces and to the highest vibrational and acoustic levels. These speeds are associated with resonant excitation phenomena (conventional or parametric resonance) of certain natural modes of the gearbox.

There are many sources of vibrational excitation for a geared system [1, 2, 3, 4]. For a gearbox operating at a constant average speed, the excitations and, in particular, the static transmission error, are periodical functions. As it is possible to fix the operating speed of a gearbox over a wide range of speeds, each of the harmonic components of the static transmission error has the capacity to excite a mode of the gearbox in a resonant manner. The excitation of some modes leads to strong amplifications of the mesh force. This type of phenomenon has been brought to light by experimental investigations [5] and by using simplified models [6]. The object of our study is to introduce a more complete dynamic model with a view to indicating the influence of each of the mechanical components of a gearbox on the localisation of critical rotational speeds.

**2- MODEL OF THE GEARBOX**

The gearbox studied is fitted out with a helical gear pair. The main characteristics of the gear are presented in Table 1. The driving and the driven wheels are identical.

| Z | 49 | Normal module | 3.5 mm |
|---|---|---|---|
| **Base radius (mm)** | 80.5 mm | **Facewidth** | 35 mm |
| **Transverse pressure angle $\alpha$** | 20° | **Centre distance** | 171.5 mm |
| **Base helix angle $\beta$** | 20° | **Total contact ratio $e_\gamma$** | 2.878 |

*Table1. Gear characteristics.*

The shafts and the casing of the gearbox were discretised using a finite element method. The shafts were modelled using beam elements with two nodes and 6 degrees of freedom for each node (see Figure 1).

Each toothed wheel was modelled with concentrated mass elements and rotary inertias at three nodes along the shaft.

The contact zone between the teeth was modelled using a symmetrical 12x12 stiffness matrix which couples the 6 degrees of freedom of the driven wheel with the 6 degrees of freedom of the driving wheel. This matrix is defined from the geometrical characteristics of the gear and from the mesh stiffness, according to the method described in [5].

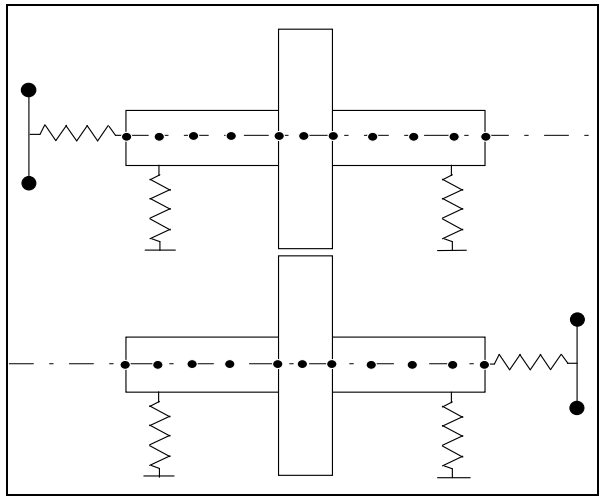

*Figure 1. Model of the shafts.*

The input and output inertias corresponding to the motor and the load are equal to 1 kg.m². They are connected to the shafts through torsion stiffness elements which model the elasticity of the couplings utilised to connect the motor and the load to the gearbox.

The elastic coupling induced by the rolling element bearings between the shafts and the casing is complex. On the one hand, their stiffnesses are non-linear functions of the transmitted load. On the other hand, they induce elastic couplings between 5 degrees of freedom of the shaft and 5 degrees of freedom of the casing [7]. Each bearing can be modelled with a 10x10 stiffness matrix, after linearising each term around the static equilibrium position. In this study, we supposed that the coupling terms were negligible. For each of the four identical bearings in the gearbox, we introduced one axial stiffness, two radial stiffnesses situated in two perpendicular directions, and two stiffnesses coupling the rotation of the free sections along the shafts to the box (rotational stiffnesses).

The steel casing is a 450x280x160 mm parallelepiped which is 10 mm thick. it was modelled using shell elements and three-dimensional structural solids to represent the bearing housings (see Figure 2).

The elastic model of the whole gearbox has approximately 1000 elements and 7000 degrees of freedom.

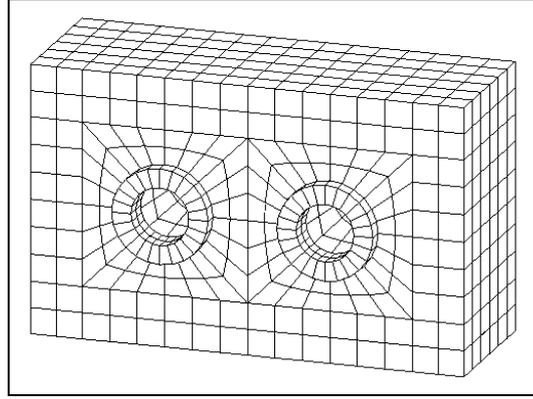

*Figure 2. Model of the casing.*

### 3- COMPUTATION OF THE VIBRATORY RESPONSE

In studying the forced vibrations of the gearbox, we supposed that the only excitation is the static transmission error [2]. The main contributions to this error are the fluctuation of the mesh stiffness and the geometry faults of the gear pair [1, 8, 9]. In the stationary regime, they induce a periodical excitation. The matrix equation which governs vibrations of the discretized gearbox can be written as follows :

$$M\ddot{x} + Kx + k(t)Dx = F + E(t) \qquad (1)$$

In this equation, **M** and **K** are the classical mass and stiffness matrices provided by the finite element method and **F** is the classical external force vector. Matrix **D** is derived from the geometric characteristics of the gear pair, **E** is an equivalent force vector induced by the unloaded static transmission error and **k(t)** is the periodic mesh stiffness. In the modal base defined from the time-invariant homogeneous counterpart equation and by introducing a viscous modal damping, the matrix equation (1) can be changed into :

$$m\ddot{q} + c\dot{q} + kq + g(t)dq = s \qquad (2)$$

Here, **m**, **c** and **k** are the mass, damping and stiffness diagonal matrices, **s** is the modal force vector and **g(t).d** is a non-diagonal matrix which is induced by the fluctuating part of the mesh stiffness and which couples the equations of motion. To solve the equation (2), we used the Spectral Iterative Method described in [5]. This method provides a direct spectral description of the vibratory response at every degree of freedom.

The object of this study is to characterise the influence of different mechanical parameters on the modes which lead to high dynamic mesh force. We chose a static transmission error resulting from a harmonic variation of the mesh stiffness. We chose an average value of the mesh stiffness equal to $4.10^8$ N/m and a peak amplitude of its harmonic fluctuation equal to $4.10^7$ N/m. The constant motor torque was equal to 500 Nm. We chose equivalent viscous damping rates of 3% for each mode.

## 4- MODES EXCITED BY THE STATIC TRANSMISSION ERROR

It is possible to identify the modes which can be excited by the static transmission error using an energy approach. For the j-th mode, the local potential energy $U_m^{(j)}$ associated with the mesh stiffness, the total potential energy $U_T^{(j)}$ and the energy rate $\rho_m^{(j)}$ are defined by:

$$U_m^{(j)} = \frac{1}{2}({}^t\phi^{(j)}[k_m]\phi^{(j)})$$

$$U_T^{(j)} = \frac{1}{2}({}^t\phi^{(j)}[K_T]\phi^{(j)})$$

$$\rho_m^{(j)} = \frac{U_m^{(j)}}{U_T^{(j)}}$$

where **[$k_m$]** is the local stiffness matrix associated with the mesh stiffness and **[$K_T$]** is the global stiffness matrix.

The modes which have the highest $\rho_m$ energy rate are called the $\phi_m$ modes. The higher $\rho_m$ is, the higher the dynamic mesh load should be.

For each bearing, the $\rho_{ra}$ energy rates associated with the radial stiffness elements and the $\rho_{ro}$ energy rates associated with the rotational stiffness elements can similarly be defined.

## 5- INFLUENCE OF SHAFT BENDING

Our first step was to calculate the natural modes of the gearbox for a model which only takes into account torsion vibrations of shafts, gear and couplings (see Figure 3a). By analysing the mode shapes, two $\phi_m$ modes with $\rho_m$ energy rates equal to 34% (1531 Hz) and 61% (3140 Hz) could be detected. Figure 3b displays the evolution of the root mean square value of the mesh force relative to the mesh frequency. The resonances of the $\phi_m$ modes generate the highest mesh loads.

We then calculated the natural modes for the gearbox using a model taking into account tension, compression, torsion and bending vibrations of shafts and gear (see Figure 4a). The casing and the bearings were assumed to be rigid. The shafts were considered to be simply supported. The presence of three $\phi_m$ modes with $\rho_m$ energy rates equal to 22%, 26% and 27% is noted (1251 Hz, 3056 Hz and 3682 Hz). Unlike the other modes of the structure, the shapes of these modes simultaneously offer bending and torsion vibrations. This phenomenon is explained by the fact that the gear couples the complete set of degrees of freedom of the driven wheel to the complete set of degrees of freedom of the driving wheel. Figure 4b displays the evolution of the root mean square value of the mesh force relative to the mesh frequency. The $\phi_m$ modes generate three resonances. The difference between these results and the previous ones shows a shaft bending effect. A torsion model does not allow the appropriate prediction of the critical rotational speeds. It is further noted that this type of model over-estimates the level of dynamic overloads.

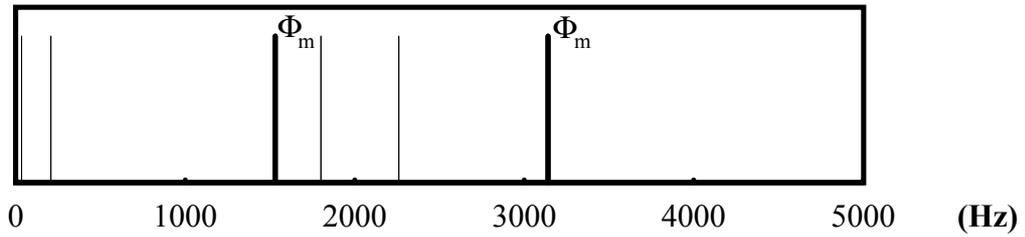

Figure 3a. Natural frequencies of the model including torsion vibrations of shafts, gear and couplings.

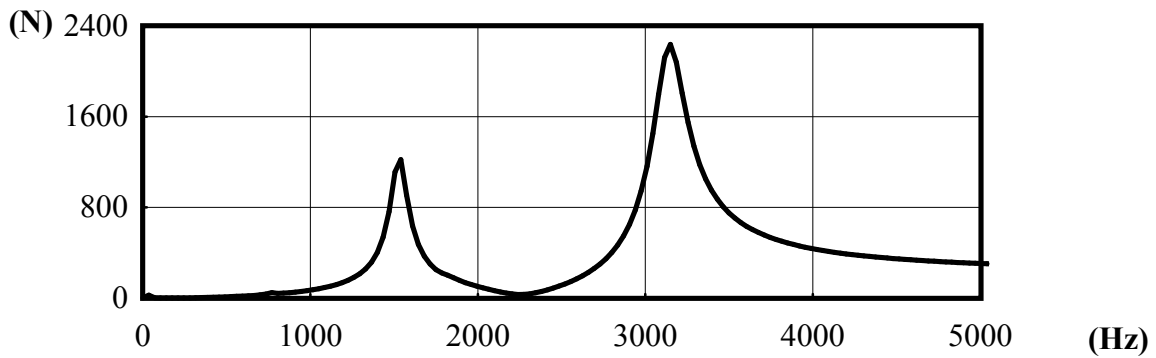

Figure 3b. Root mean square value of the mesh force versus mesh frequency.

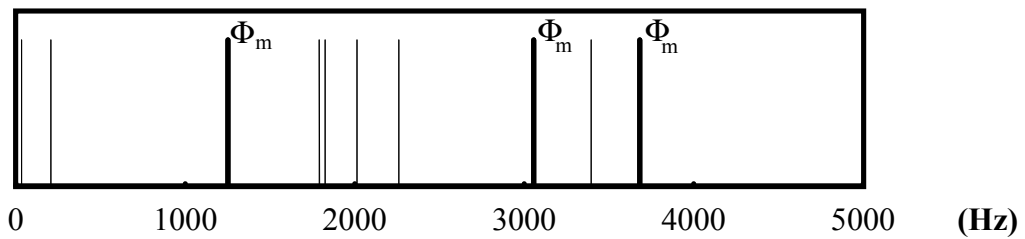

Figure 4a. Natural frequencies of the model including torsion, tension, compression and bending vibrations of shafts and gear.

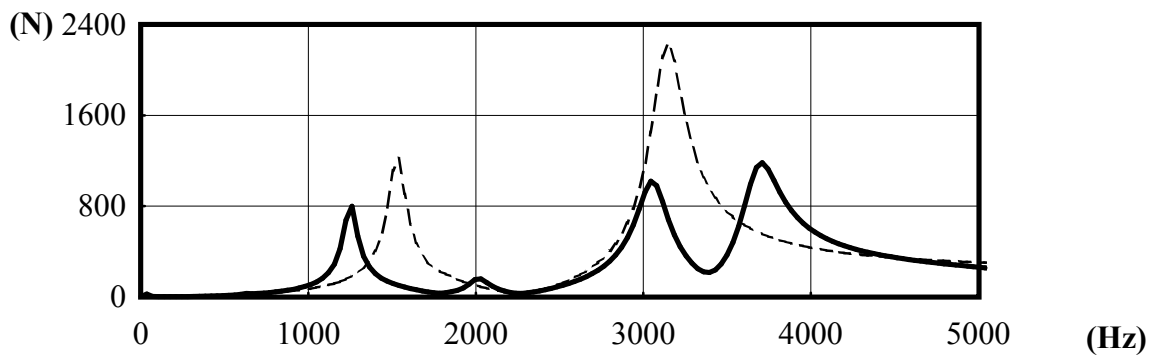

Figure 4b. Root mean square value of the mesh force versus mesh frequency. (Torsion, tension, compression and bending————; Torsion --------).

## 6- INFLUENCE OF THE ROLLING ELEMENT BEARINGS

The numerical results presented below were obtained from a model which takes into account not only the tension, compression, torsion and bending vibrations of shafts and gear, but also the elasticity of bearings. The nominal values of the radial (identical in the two perpendicular directions), axial and rotational stiffnesses of the bearings were respectively equal to $10^9$ N/m, $10^8$ N/m and $10^6$ Nm/rad.

Figure 5a supplies the first natural frequencies associated with this new model. The first and the second natural frequencies of the structure (16 Hz and 95 Hz) correspond to pure torsion modes. They are controlled by couplings. The gearbox presents two axial modes (590 Hz and 605 Hz), four bending modes (1728 Hz, 2237 Hz, 2898 Hz, 2961 Hz), and one torsion mode for the shafts (1768 Hz). The $\phi_m$ modes (1191 Hz, 3365 Hz and 4157 Hz) are not the first modes of the system.

The figure 5b displays the evolution of the root mean square value of the mesh force relative to the mesh frequency. The presence of three resonances is noted. Comparison with the rigid bearings model shows that the elasticity of the bearings modifies the critical rotational speeds.

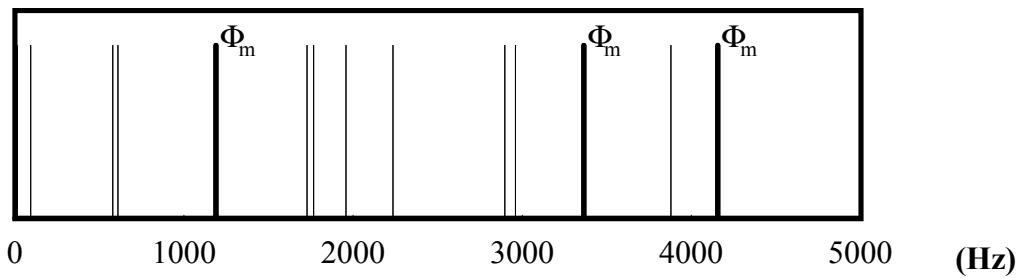

Figure 5a. Natural fequencies of the model including the vibration of the bearings .

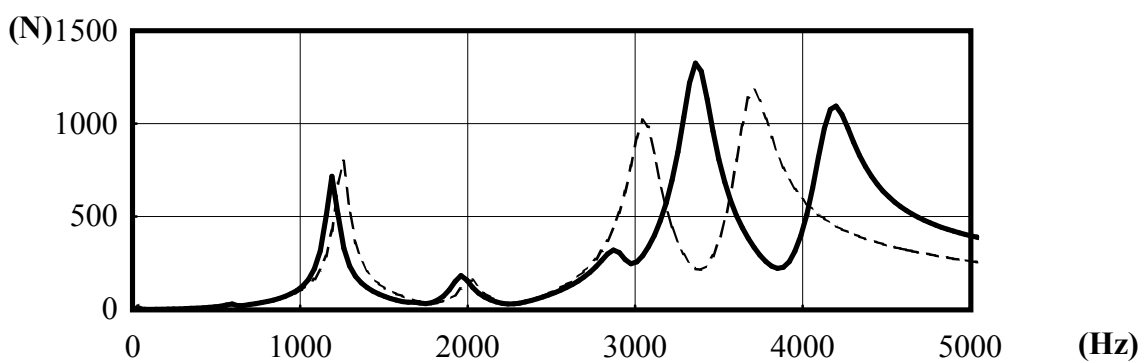

Figure 5b. Root mean square value of the mesh force versus mesh frequency.
(Elastic bearings ⎯⎯⎯; Rigid bearings --------).

### Influence of radial stiffness

In order to analyse in more detail the influence of the rolling element bearings on modes which lead to a high mesh force, we have varied the radial stiffness within the range $10^8$ N/m to $10^{10}$ N/m. This range corresponds to the usual values encountered in geared systems. By comparison, the average mesh stiffness value for the gearbox studied is $4 \cdot 10^8$ N/m.

Figure 6 displays the map of the root mean square value of the mesh force relative to radial stiffnesses of bearings and mesh frequency. For the gearbox studied and within the considered stiffness variation range ($10^8$ N/m to $10^{10}$ N/m), we observed the following characteristics :

1- The frequencies of the $\phi_m$ modes change with the radial stiffnesses of the bearings. This evolution appears for radial stiffnesses which are all the greater when the frequency of the $\phi_m$ mode is high. It depends on the modal potential energy associated with the radial stiffness elements, that is the $\rho_{ra}$ energy rates of the $\phi_m$ modes. Thus:

- For the first $\phi_m$ modes (frequency lower than 2000 Hz), the $\rho_{ra}$ energy rates are high between $10^8$ and $10^9$ N/m and then lower from $10^9$ N/m. So, the evolution of the frequencies of the $\phi_m$ modes is high between $10^8$ and $10^9$ N/m, and then weak from $10^9$ N/m.

- For the $\phi_m$ modes situated at higher frequencies (frequency over 2000 Hz), the $\rho_{ra}$ energy rates are nil between $10^8$ and $10^9$ N/m and increase from $10^9$ N/m. The evolution of the frequencies of the $\phi_m$ modes is initially fairly weak between $10^8$ and $10^9$ N/m, and then more obvious from $10^9$ N/m.

2- The evolution of the frequency of the $\phi_m$ modes is accompanied by a modification of the mode shape so that the $\rho_m$ energy rates of the $\phi_m$ modes, that is the potential energy associated with the mesh stiffness, decrease.

3- The resonances of the $\phi_m$ modes (1676 Hz, 3349 Hz and 4139 Hz for K=$10^8$ N/m), in the end disappear to the benefit of the onset of new $\phi_m$' modes situated at lower frequencies (respectively 1364 Hz, 3311 Hz and 4053 Hz for K= $10^{10}$ N/m). This behaviour leads to lower critical rotational speeds when the radial stiffnesses are greater.

### Influence of rotational stiffness

We also varied the rotational stiffness of the bearings between $10^4$ Nm/rad and $10^8$ Nm/rad. This range enables the study of the influence of the different types of bearings : ball-type bearings (low rotational stiffness), cylindrical roller-type bearings (average rotational stiffness) and taper roller-type bearings (high rotational stiffness).

Figure 7 displays the map of the root mean square value of the mesh force relative to rotational stiffnesses of bearings and mesh frequency. The frequencies of the $\phi_m$ modes change with the rotational stiffness of the bearings. The evolution of these frequencies depends on the $\rho_{ro}$ energy rates associated with the rotational stiffness elements. The $\rho_{ro}$ energy rates are high between $10^6$ and $10^7$ Nm/rad. So, the evolution of the frequencies of the $\phi_m$ modes is high between $10^6$ and $10^7$ Nm/rad. The resonances of the $\phi_m$ modes (respectively 1110 Hz, 3100 Hz and 3914 Hz for K=$10^4$ Nm/rad) disappear to the benefit of the onset of new $\phi_{m'}$ modes (respectively 1259 Hz and 3459 Hz for K=$10^8$ Nm/rad).

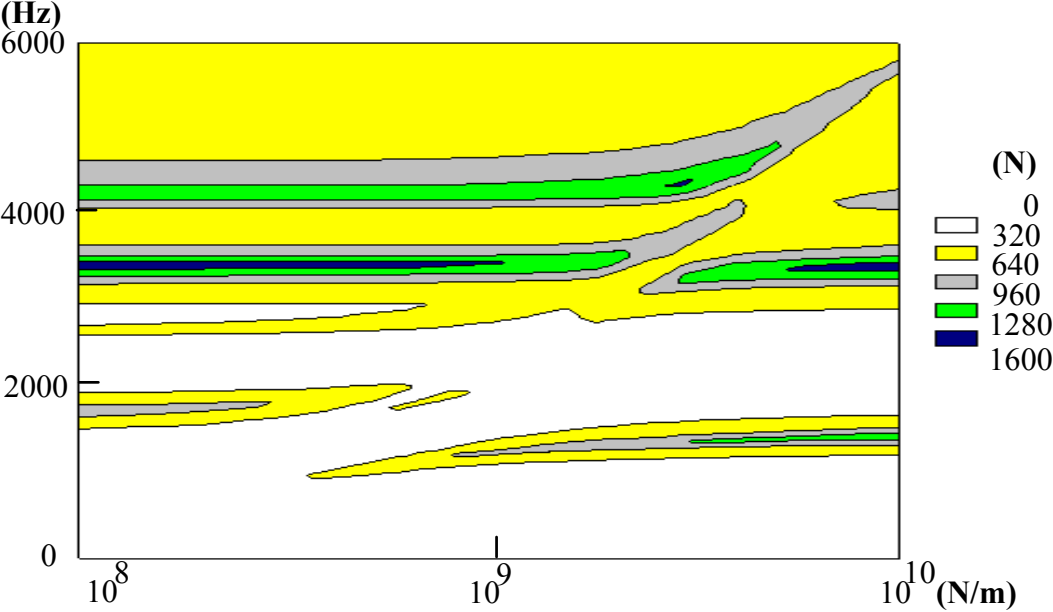

Figure 6. Root mean square value of the mesh force versus radial stiffnesses of bearings (X axis) and mesh frequency (Y axis).

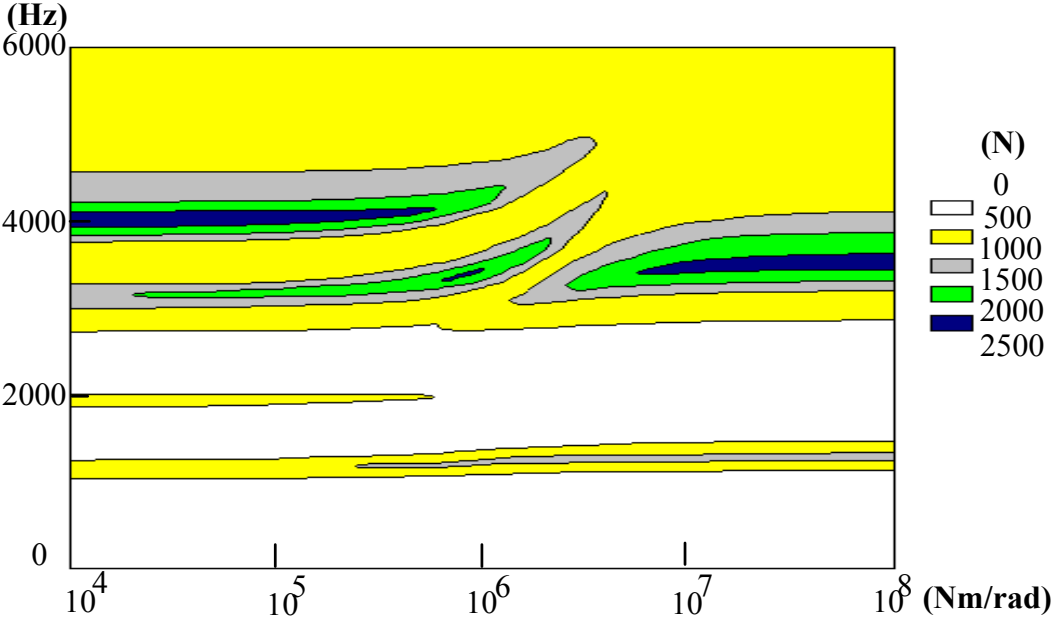

Figure 7. Root mean square value of the mesh force versus rotational stiffnesses of bearings (X axis) and mesh frequency (Y axis).

## 7- INFLUENCE OF THE CASING

In order to assess the influence of the mechanical properties (elasticity and inertia) of the casing on the critical rotational speeds of the gearbox, a modal analysis of the complete gearbox was carried out. The model takes account of the whole set of components of the gearbox, that is to say the gear, the shafts, the bearings, the casing, the couplings and the input and output inertias. For this new study, we chose radial stiffnesses of bearings equal to $10^9$ N/m and rotational stiffnesses equal to $10^6$ Nm/rad. Figure 8 supplies the first natural frequencies of the new model.

A detailed examination of the new modes enabled us to conclude that :
- Within the considered frequency range, the $\phi_m$ modes are numerous but the $\rho_m$ energy rate in each is fairly low.
- The vibrations of the shafts are coupled with the vibrations of the casing. The dynamic behaviour of the casing thus interacts with the gear set.
- Because of vibrations of the casing, the frequency and the shape of each $\phi_m$ mode are different from those calculated with a rigid casing.

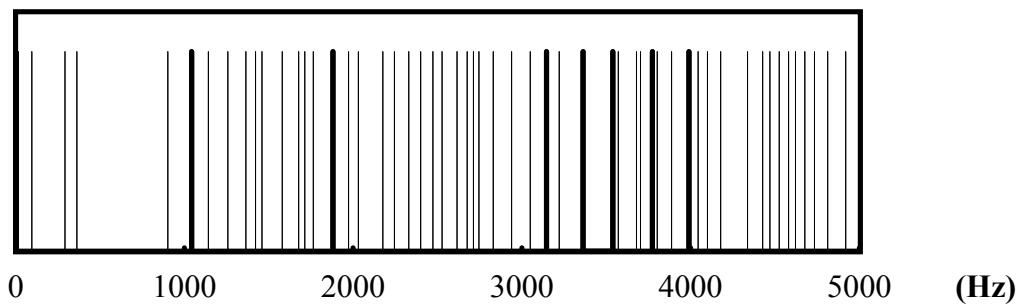

Figure 8. Natural frequencies of the system. Model including casing elasticity.

We then compared the evolution of the root mean square value of the mesh force with the previous results (rigid casing). This comparison is offered for radial stiffnesses-rotational stiffnesses of bearings respectively equal to $10^8$ N/m-$10^4$ Nm/rad (Figure 9a), $10^9$ N/m-$10^6$ Nm/rad (Figure 9b) and $10^{10}$ N/m-$10^8$ Nm/rad (Figure 9c).

As displayed in Figure 9a, the mechanical properties of the casing do not change the $\phi_m$ modes for radial stiffnesses equal to $10^8$ N/m and rotational stiffnesses equal to $10^4$ Nm/rad. In point of fact, the casing vibrations are not coupled with those of the gear. Conversely, for higher bearing stiffnesses, the dynamic behaviour of the gear is then highly affected by the elastic properties of the casing. As illustrated in Figures 9b and 9c, the resonances are more numerous. The frequencies of these resonances have changed appreciably and the maximum level of the mesh force has dropped.

Parametric studies have shown that rotational stiffnesses play a more significant role than the radial stiffnesses of the coupling between the casing and the gear set.

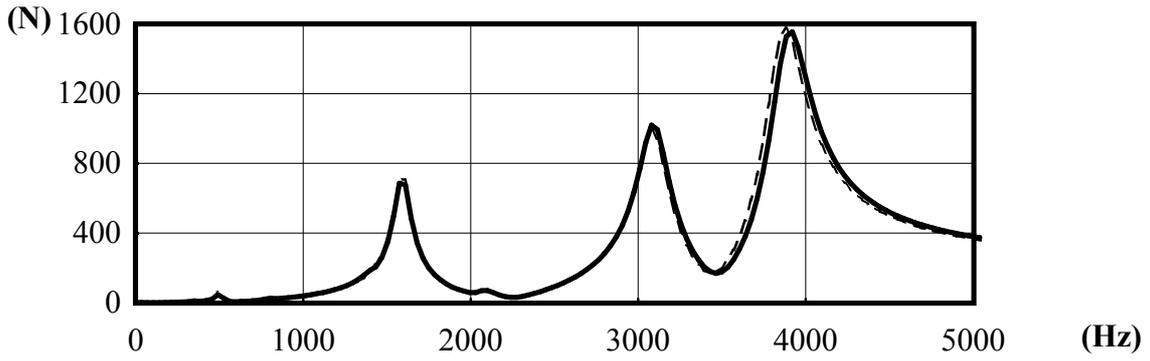

*Figure 9a. Root mean square value of mesh force versus mesh frequency.
Elastic casing (―――――) and rigid casing (― ― ― ―).
Krot = $10^4$ Nm/rad. Krad = $10^8$ N/m.*

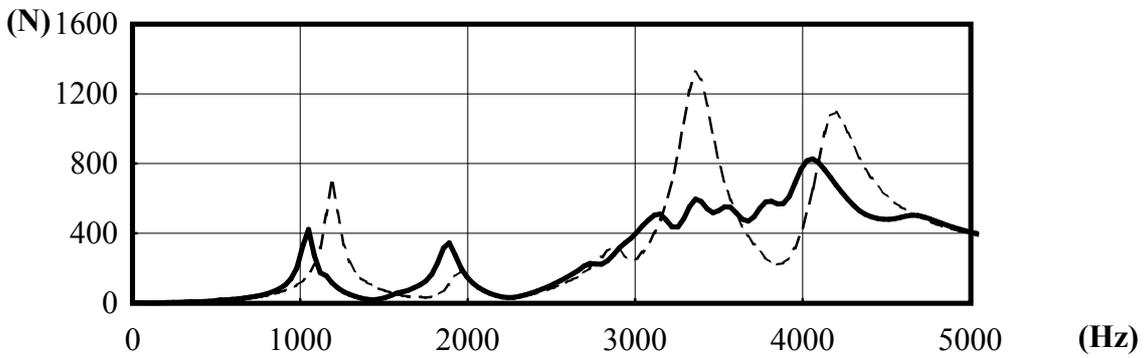

*Figure 9b. Root mean square value of the mesh force versus mesh frequency.
Elastic casing (―――――) and rigid casing (― ― ― ―).
Krot = $10^6$ Nm/rad. Krad = $10^9$ N/m.*

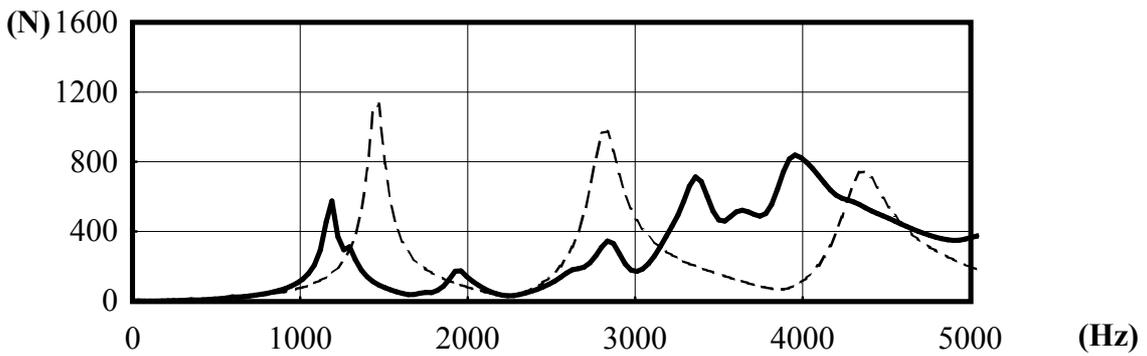

*Figure 9c. Root mean square value of the mesh force versus mesh frequency.
Elastic casing (―――――) and rigid casing (― ― ― ―).
Krot = $10^8$ Nm/rad. Krad = $10^{10}$ N/m.*

## 8- INFLUENCE OF THE TORSION STIFFNESS OF THE FLEXIBLE COUPLINGS

We have connected the torsion stiffness of the coupling and the torsion stiffness of the adjacent half-shaft in series. The figures 10a and 10b show that, for a low torsion stiffness of the coupling, the equivalent stiffness of the coupling+shaft set is comparable to the stiffness of the coupling. For a torsion stiffness of the coupling which is greater than that of the shaft, the torsion stiffness of the coupling+shaft set converges towards the stiffness of the shaft.

The evolution of the torsion stiffness of the coupling+shaft set is significantly different according to whether the shafts are long (and thus flexible) or short (and thus stiff). For 160 mm-long shafts, the coupling+shaft set has a stiffness which constantly remains below the torsion stiffness of the gear set. For 80-mm long shafts, the sitffness of the coupling+shaft set is initially below the torsion stiffness of the gear set. It becomes equal to and then greater than this stiffness when the torsion stiffness of the couplings increases.

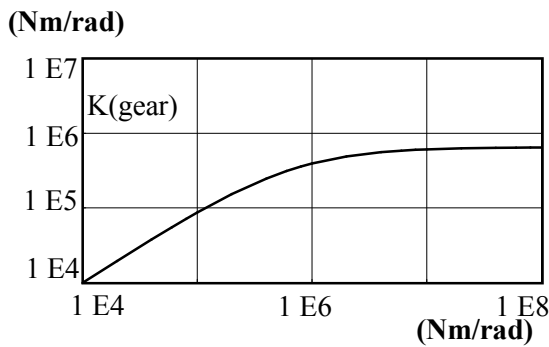 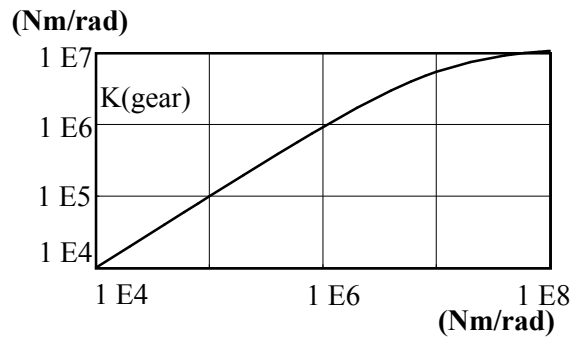

*Figure 10a. Long shafts.*          *Figure 10b. Short shafts.*

*Torsion stiffness of the coupling+shaft set versus torsion stiffness of the coupling.*

Figures 11 and 12 display the evolution of the root mean square value of the mesh force for a flexible coupling ($10^4$ Nm/rad) and a stiff coupling ($10^8$ Nm/rad). The model used is the model where the gearbox casing is rigid.

For long shafts, we see that a variation of the stiffness of the couplings only modifies the first torsion modes of the gearbox. The $\phi_m$ modes (3350 Hz and 4155 Hz) are identical for flexible couplings and stiff couplings. Any variation of the torsion stiffness of the couplings thus has no appreciable influence on the critical rotational speeds.

For short shafts, the frequency of the $\phi_m$ modes is equal to 1388 Hz and 2717 Hz when the torsion stiffness of the couplings is low ($10^4$ Nm/rad) and it is equal to 208 Hz and 3427 Hz when the torsion stiffness of the couplings is greater ($10^8$ Nm/rad). These differences translate the appreciable influence of the torsion stiffness of the couplings on the dynamic behaviour of the gearbox. In point of fact, the frequency of the $\phi_m$ modes increased with the stiffness of the couplings (from 1388 to 2084 Hz and from 2717 to 3427

Hz) while the amplitude of the resonance decreased. New $\phi_m'$ modes appeared at lower frequencies (208 Hz and 661 Hz).

For a coupling stiffness equal to $10^4$ Nm/rad, the motor and the load correspond to vibration nodes of the $\phi_m$ modes. They are totally disconnected from the gear set by the distortion of the couplings. For a stiffness equal to $10^8$ Nm/rad, the first $\phi_m$ mode (208 Hz) corresponds to the first mode of the gearbox. This mode is dominated by torsion vibrations. The shafts, the couplings, the motor and the load move as rigid bodies. The gear set being the most flexible torsion element, this alone is distorted. For short shafts and stiff couplings, a $\phi_m$ mode with a low frequency, which is related to the presence of high input and output inertias, thus appears.

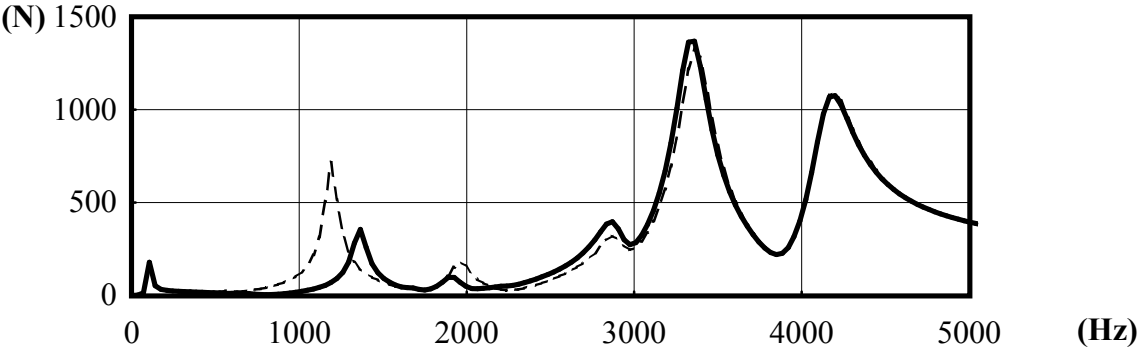

Figure 11. Long shafts. Root mean square value of the mesh force versus mesh frequency.
Kacc = $10^4$Nm/rad ( ------- ); Kacc = $10^8$Nm/rad (———).

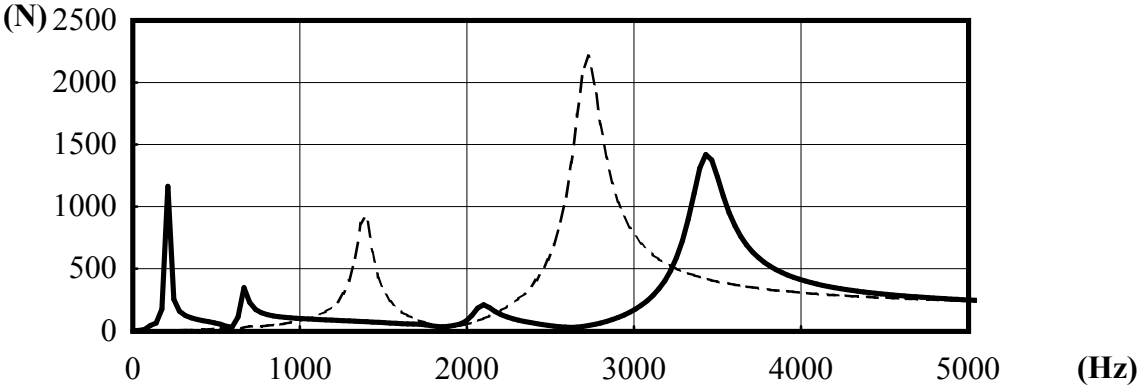

Figure 12. Short shafts. Root mean square value of the mesh force versus mesh frequency.
Kacc = $10^4$Nm/rad ( ------- ); Kacc = $10^8$Nm/rad (———).

## 9- CONCLUSION

To study the dynamic behaviour of a helical gearbox we modelled its components using the finite element method.

We defined a method enabling the identification of the $\phi_m$ modes which can be excited by the static transmission error, based on a modal analysis of the gearbox.

We showed that these $\phi_m$ modes generated the highest dynamic mesh forces and consequently the strongest vibration levels of the casing.

We showed the influence of bending along the shafts, of the elasticity of the bearings, of the mechanical properties of the casing and of the torsion stiffnesses of the couplings on the critical rotational speeds of the gearbox.

We can thus conclude that the critical rotational speeds for a geared system can only be realistically predicted by a global model including the shafts, the bearings, the casing, the couplings and the input and output inertias.